# Effect of Physical Inactivity on the Oxidation of Saturated and Monounsaturated Dietary Fatty Acids: Results of a Randomized Trial

Audrey Bergouignan[1], Dale A. Schoeller[2], Sylvie Normand[3], Guillemette Gauquelin-Koch[4], Martine Laville[3], Timothy Shriver[2], Michel Desage[3], Yvon Le Maho[1], Hiroshi Ohshima[5], Claude Gharib[6], Stéphane Blanc[1]*

1 Institut Pluridisciplinaire Hubert Curien, Département d'Ecologie, Physiologie, et Ethologie, Strasbourg, France, 2 Department of Nutritional Sciences, University of Wisconsin–Madison, Madison, Wisconsin, United States of America, 3 Centre de Recherche en Nutrition Humaine de Lyon, Faculté de Médecine Laennec, Lyon, France, 4 Centre National d'Etudes Spatiales, Paris, France, 5 Japan Aerospace Exploration Agency, Tsukuba Space Center, Tsukuba, Ibaraki, Japan, 6 Laboratoire de Physiologie de l'Environnement, Faculté de Médecine Grange Blanche, Lyon, France

**Trial Registration:** ClinicalTrials.gov: NCT00311571

**Funding:** Our participation in the study was supported by NIH grant RO1 DK 30031 (Dale A. Schoeller) and CNES grants (Stéphane Blanc and Claude Gharib). Stéphane Blanc was a fellow of the Association Française de Nutrition and the Société de Nutrition de Langue Française. The sponsors played no role in the collection, analysis, interpretation of data, writing of the paper or the decision to submit it for publication.

**Competing Interests:** The authors have declared that no competing interests exist.

**Citation:** Bergouignan A, Schoeller DA, Normand S, Gauquelin-Koch G, Laville M, et al. (2006) Effect of physical inactivity on the oxidation of saturated and monounsaturated dietary fatty acids: Results of a randomized trial. PLoS Clin Trials 1(5): e27. DOI: 10.1371/journal.pctr. 0010027

**Received:** April 6, 2006
**Accepted:** August 16, 2006
**Published:** September 29, 2006

**DOI:** 10.1371/journal.pctr.0010027

**Abbreviations:** CI, confidence interval; DIT, diet-induced thermogenesis; FFM, fat-free mass; FM, fat mass; LPL, lipoprotein lipase; MEDES, Institut de Médecine et Physiologie Spatiale (Institute for Space Physiology and Medicine); NPRQ, nonprotein respiratory quotient; RMR, resting metabolic rate; SD, standard deviation; TG, triglyceride

* To whom correspondence should be addressed. E-mail: stephane. blanc@c-strasbourg.fr

## ABSTRACT

**Objectives:** Changes in the way dietary fat is metabolized can be considered causative in obesity. The role of sedentary behavior in this defect has not been determined. We hypothesized that physical inactivity partitions dietary fats toward storage and that a resistance exercise training program mitigates storage.

**Design:** We used bed rest, with randomization to resistance training, as a model of physical inactivity.

**Setting:** The trial took place at the Space Clinic (Toulouse, France).

**Participants:** A total of 18 healthy male volunteers, of mean age ± standard deviation 32.6 ± 4.0 y and body mass index 23.6 ± 0.7 kg/m$^2$, were enrolled.

**Interventions:** An initial 15 d of baseline data collection were followed by 3 mo of strict bed-rest alone (control group, $n = 9$) or with the addition of supine resistance exercise training every 3 d (exercise group, $n = 9$).

**Outcome measures:** Oxidation of labeled [d$_{31}$]palmitate (the main saturated fatty acid of human diet) and [1-$^{13}$C]oleate (the main monounsaturated fatty acid), body composition, net substrate use, and plasma hormones and metabolites were measured.

**Results:** Between-group comparisons showed that exercise training did not affect oxidation of both oleate (mean difference 5.6%; 95% confidence interval [95% CI], −3.3% to 14.5%; $p = 0.20$) and palmitate (mean difference −0.2%; 95% CI, −4.1% to 3.6%; $p = 0.89$). Within-group comparisons, however, showed that inactivity changed oxidation of palmitate in the control group by −11.0% (95% CI, −19.0% to −2.9%; $p = 0.01$) and in the exercise group by −11.3% (95% CI, −18.4% to −4.2%; $p = 0.008$). In contrast, bed rest did not significantly affect oleate oxidation within groups. In the control group, the mean difference in oleate oxidation was 3.2% (95% CI, −4.2% to 10.5%; $p = 0.34$) and 6.8% (95% CI, −1.2% to 14.7%; $p = 0.08$) in the exercise group.

**Conclusions:** Independent of changes in energy balance (intake and/or output), physical inactivity decreased the oxidation of saturated but not monounsaturated dietary fat. The effect is apparently not compensated by resistance exercise training. These results suggest that Mediterranean diets should be recommended in sedentary subjects and recumbent patients.

**Editorial Commentary**

**Background:** Obesity is an important contributor to the burden of chronic diseases, particularly type II diabetes, cardiovascular disease, hypertension, and stroke. Being inactive is a risk factor for all of these conditions. However, the physiological effects of inactivity are not well understood. In this trial, supported by the European Space Agency, a group of researchers aimed to further understand the effects of physical inactivity on the way that fat from the diet is metabolized (i.e., broken down to generate energy). 18 healthy male volunteers were randomized into two groups, both of whom underwent 90 days of bed rest, aiming to mimic sedentary behavior. One group also received an exercise training program during the 90 days' bed rest. The researchers examined to what extent two different types of fatty acids common in the diet were metabolized over the duration of the trial: oleate (monounsaturated fat) and palmitate (saturated fat). As secondary objectives of the study, body weight, water, fat, and energy expenditure were also examined in the participants.

**What this trial shows:** The researchers did not see any statistically significant changes between the groups—that is, participants receiving bed rest, and those receiving bed rest plus exercise training—for any of the primary or secondary outcomes, except for resting metabolic rate, which was higher in the exercise group. However, they did see physiologically relevant changes in fat metabolism of one of the fatty acids, palmitate, over the course of the trial within both groups studied. Although metabolism of oleate (monounsaturated fat) did not show significant changes over the course of the trial, metabolism of palmitate (saturated fat) dropped by nearly 10% in both groups (bed rest, and bed rest plus exercise).

**Strengths and limitations:** The study design was appropriate to the questions being posed, and the techniques for examining fat metabolism were relevant. Although the number of participants was very small, this problem is true of many such studies due to the cost and complexity of the interventions. The model for inactivity used in this trial—90 days' bed rest—is very extreme. Very few studies of this type have been performed, with most of the evidence relating to activity and fat handling coming from training studies in otherwise sedentary people.

**Contribution to the evidence:** It is already known that physical activity has numerous health benefits, including the prevention of obesity. This trial provides data showing that inactivity lowers the ability to metabolize fat, specifically saturated fat, from the diet, which would therefore be more likely to be stored in the body.

*The Editorial Commentary is written by PLoS staff, based on the reports of the academic editors and peer reviewers.*

# INTRODUCTION

Obesity is reaching pandemic proportions, affecting all sexes, races, and ages. Currently more than 1 billion adults are overweight and at least 300 million of them are clinically obese. Overweight, and ultimately obesity, is a major contributor to the global burden of chronic diseases and disabilities such as type 2 diabetes, cardiovascular diseases, hypertension, and stroke, and certain forms of cancer [1]. As a consequence, obesity accounts for 2%–6% of total health care costs in several developed countries.

Obesity is a fat storage disease. In humans, stored lipids originate largely from the diet. Obesity thus represents a failure in dietary fat balance. Using [1,1,1,-$^{13}$C]triolein, a long-chain triacylglycerol, Binnert et al. [2] showed that dietary fat oxidation is decreased by 50% in obese women as compared to lean counterparts. Furthermore, it has been reported that there is no improvement in such preferential dietary fat channeling away from oxidation amongst post-obese weight-stable women [3], lending support to the hypothesis that the partitioning of dietary fat to storage is a causal factor in obesity rather than an adaptive response to obesity. Studies in obese Zucker rats has shown that this impaired partitioning between oxidation and storage is likely due to a preferential trafficking of meal-derived fatty acids towards adipose tissue for storage [4]. Understanding the factors which regulate dietary fat oxidation is therefore crucial to deciphering the causes of obesity.

We hypothesize that the generalized sedentary behaviors of our modern societies, which have been increasing since the beginning of the last century [5], play a key role in the reduced capacity to use dietary fat as fuel. It is already well accepted that physical inactivity on its own represents a risk factor for numerous chronic diseases including obesity and, as such, has been classified as the second cause of death in the US [6]. Nevertheless, it is also important to note that our knowledge of the detrimental health effects of inactivity is somewhat indirect and based on the positive effects of exercise training on sedentary populations [7]. Because there are very few longitudinal studies of increasing sedentarianism, our knowledge of the development of the "sedentary death syndrome" as proposed by Lees and Booth remains weak [8].

We investigated the effects of enforced physical inactivity induced by three months of strict bed rest on dietary fat oxidation and tested the efficacy of a low-volume, high-intensity resistance training program to mitigate the effects of this extreme sedentary behavior. This paper reports the primary outcomes investigating nutrition. Data on other physiological functions in this inactivity regimen are already published [9–16] or are to appear elsewhere.

# METHODS

## Participants

A call for candidates was made via the Internet on the MEDES (Institut de Médecine et Physiologie Spatiale, Toulouse, France) and ESA (European Space Agency) Web sites and by press announcements. The selection was carried out in two phases.

**Preselection: First screening.** Preselection was based on the volunteers' application files, comprising first a general questionnaire on the participant's way of life, education, and professional experience; and second, a medical questionnaire on personal and family medical history, by phone. The purpose was to select volunteers who met the following requirement criteria: healthy, male, European Community citizen, aged 25 to 45, nonsmoker, no alcohol, no drug dependence, no medical treatment, height 165 to 185 cm, no overweight nor excessive thinness (body mass index as the ratio of weight [kg] to height [m$^2$] between 20 and 27), no personal nor family history of chronic or acute disease (e.g., hypertension, diabetes) that could affect the physiological data and/or create a risk for the participant during the experiment. In addition, any participant covered by a Social Security system was to be free of any engagement during four consecutive months.

Specific exclusion criteria were: having given blood (more than 300 ml) in a period of three months or less before the start of the experiment, participant already participating in a clinical research experiment, poor tolerance to blood sampling, past record of orthostatic intolerance, cardiac rhythm disorders, allergies, intensive sport training, fractures or tendon laceration since less than one year, chronic back pain, history of thrombophlebitis, presence of metallic implants, special dietary requirements, sleep disorders, or photosensitive epilepsy. Out of the 730 applications received, 124 participants were preselected.

**Selection: Second screening.** The 124 preselected participants were invited to MEDES to undergo medical and psychological examinations to select the participants most apt to participate in the long-duration bed-rest intervention. Tests lasted 48 hours and were carried out by physicians and psychologists from MEDES who were not involved in the individual scientific protocols. The purpose of the first step was to select 17 participants (14 participants, three replacements) for the first experimentation phase in 2001. Similarly, the second selection process started at the end of the first period. The medical checkup included: clinical examination and questionnaire to verify that no exclusion criteria were present; 12-lead electrocardiogram and measurement of arterial blood pressure and heart rate; testing for orthostatic hypotension; a questionnaire to evaluate the quality of participants' sleep; ophthalmologic examination (visual acuity and fundus examination); and biochemistry, hematology, serology, and toxicology analyses. In addition, in specialized departments (Rangueil Hospital, Toulouse, France) the following tests were conducted: echo-Doppler measurements of the lower limbs (to eliminate participants with venous insufficiency), front and side chest radiography, a dental panoramic radiography, abdomen radiography (to ensure the absence of lithiasis), abdomen echography, DEXA (dual-energy-X-ray absorptiometry) measurement of bone density (which should be no more than 1 standard deviation [SD] above or below the age/sex-matched mean), and a measurement of maximal oxygen consumption. The preselected participants underwent psychiatric and psychological evaluations to ensure that participants would tolerate the conditions of the experiment. The selection protocol included psychological tests applied to the candidates in a group, a projective test, and an interview.

**Participant aptitude.** Candidates were declared able to participate in the experiment if the two following conditions were fulfilled: first, that the informed consent form had been carefully read and that candidates had the opportunity to ask any additional questions concerning the research project. After having given satisfactory answers to these questions, participants signed an Information and Consent Form that was approved by the Institutional Review Board of Midi-Pyrénées I (France). Second, after the results of the numerous medical and psychological tests were checked, a last verification was made to ensure that all inclusion criteria were present and that no exclusion criteria existed. Out of the 124 preselected participants, 15 declined to participate and 85 were excluded for medical and psychological reasons. Thus, 24 participants were included.

### Interventions

The intervention on activity lasted for 4 mo and was divided into three periods: a 15-d ambulatory control period, 90 d of bed rest in a head-down tilt position (−6°), and a 15-d recovery period. During the control and recovery periods, the participants were confined to MEDES. During the bed rest, all activities of daily living (i.e., showering, restroom needs, eating, etc.) were performed in supine position. Standing or seating positions were forbidden. Cameras were used at night to ensure compliance of the participants to these conditions. During bed rest, the subjects were divided into three groups: a control group that remained in bed ($n = 9$), an exercise group subjected to a supine resistance exercise training protocol concomitantly with the bed rest ($n = 9$), and a group receiving a single intravenous infusion of bisphosphonate (Pamidronate) to prevent bone demineralization ($n = 6$). Interventions on this latter group will not be further described as it was not part of the present protocol. Details can be found elsewhere [15].

**Training program.** The resistance training program was performed on the flywheel ergometer [17]. This device allows participants to perform maximal concentric and eccentric actions in the supine squat and calf press. The training sessions were programmed every 3 d during the bed rest and lasted 35 min. Progressive warm-ups preceded four sets of seven maximal concentric and eccentric repetitions in the squat, followed by four sets of 14 repetitions in the calf press. Periods of rest lasting 2 min were allowed between sets and 5 min between exercises. The energy expenditure of a session was on average 640 kJ. Such training prevents lean body mass loss and mitigates muscle strength loss while not significantly impacting total daily energy expenditure during the entire bed-rest period (theoretical impact of 0.03 on the physical activity level defined as the ratio of total energy expenditure to resting metabolic rate [RMR]).

### Diet

In order to dissociate the effects of physical inactivity from those of a positive energy balance induced by physical inactivity, we attempted to provide a eucaloric energy intake and thus maintain energy balance during ambulatory and bed-rest periods. However, diet monitoring represents a difficult challenge during bed rest, because body composition changes due to muscle atrophy and, as a consequence, changes in body mass do not reflect energy balance. Energy requirements were calculated as RMR times a physical activity factor of 1.4 and 1.2 during the control and bed-rest periods, respectively [18,19]. RMR was measured twice in each period to adjust intake for changes in fat-free mass (FFM). Water intake was provided at 3 l/d. Snacks and extra water were provided to the exercise group on the days of training to cover the energy cost of the session. The macronutrient composition of the diet was set at 30% fat, 15% protein, and 55% carbohydrate. Diet was supplied by the hospital kitchen and controlled by a dietician under the supervision of the investigators. The participants were asked to finish all food given. Leftovers, if any, were weighed and snacks were provided later during the day to cover the energy deficit. Macronutrients and energy intake were calculated based on the hospital recipes and Geni software (Geni, Micro 6, Nancy, France).

### Objectives

We hypothesized that physical inactivity, independent of its effects on energy balance, reduces the oxidation of dietary fat. To test this hypothesis we used a unique space physiology-derived model of strict bed rest to induce three months of physical inactivity in healthy participants. An additional objective was to test the hypothesis that a high-intensity, low-volume resistance exercise training protocol would mitigate the effect of inactivity on dietary fat oxidation.

## Outcomes

**Primary outcome measures.** Dietary fat oxidation was measured before (control period day 11) and at the end of the bed rest (bed rest day 78), 36 h after the last bout of exercise, through a combination of indirect calorimetry and a recently developed method by which fatty acids labeled by stable isotopes are mixed into a standard breakfast. Oleic (18:1) and palmitic (16:0) acids were selected for the study. These fatty acids represent the main monounsaturated and saturated fatty acids of the human diet (38% and 20% of total intake, respectively). Baseline breath and urine samples and fasting blood samples were collected. Then a breakfast representing 50% of the RMR (expressed as MJ/24 h) in energy (55% carbohydrates, 15% protein, and 30% fat) was offered to the participants. Part of this breakfast was composed of a liquid replacement meal (Boost High protein, Mead Johnson, United States) in which 15 mg/kg of $[d_{31}]$palmitic acid (>98% enriched; CIL, Andover, Massachusetts, United States) and 10 mg/kg of $[1-^{13}C]$oleic acid (>99% enriched, CIL) were homogenized at 65 °C or above (melting point of palmitate). Then, hourly breath and fresh urine samples were collected for 7 h. Breath $^{13}CO_2/^{12}CO_2$ and urinary $^2H/^1H$ ratios were measured on an isotope ratio mass spectrometer (GV Instruments, France). Recovery of $[1-^{13}C]$oleic acid was calculated as the instantaneous recovery of $^{13}C$ in expired $CO_2$ hourly sampled over the 7 h of the test and expressed as a percentage of the dose. Recovery of $[d_{31}]$palmitate was calculated as the cumulated recovery of $^2H$ in total body water hourly sampled through urine voids.

**Secondary outcome measures.** Body weight was measured daily on a special supine weighing device. Total body water was measured by hydrometry based on the isotope dilution of $H_2^{18}O$ in body water. Fat mass (FM) and FFM were measured by DEXA on QDR 4500 W (Hologic, Massy, France) scanner using the QDR System Software for version 11.2. Energy expenditure; total fat, carbohydrate, and protein oxidation rates; and the nonprotein respiratory quotient (NPRQ) and lipogenesis were calculated from hourly indirect calorimetry data. Fasting hormones and metabolites were measured on the day of the oral lipid load. Details on the methods for primary and secondary outcomes assessment can be found in Protocol S1.

## Sample Size

Eleven projects from international laboratories investigating each physiological function were selected by peer review for participation in this long-term bed rest experiment. Consequently, the selection of one particular primary outcome for sample size calculation is impossible for bed rest, and no preliminary data exist for such a long-term study. One of the most conservative changes induced by bed rest is muscle atrophy. Based on results obtained during a 42-d bed rest study [18], we observed a loss in muscle mass of 3.2 ± 2.0 kg (mean ± SD) on seven healthy men with a power of 94%. Nine subjects per group were required to detect the same changes at a power of 99%.

## Randomization

For the whole study, the participants were housed two per room. The pairing of subjects was based on thorough psychological examinations by psychologists expert in confinement situations, who made their decisions independently of the principal investigators. This procedure was considered essential to avoid conflicts between two participants to be confined together in a 25 m² room for three months. Pairs of subjects were then written on a small piece of paper, folded in an envelope, and mixed in a dark bag. Each intervention was also written on a small piece of paper, folded in an envelope, and mixed in a dark bag. An envelope containing a participant pair and an envelope containing an intervention group were randomly and sequentially selected for group allocation in the control, exercise, and bisphosphonate groups.

## Blinding

No blinding was performed in this study. Blinding was not possible given (1) the permanent inpatient design of this longitudinal study and (2) the type of countermeasures applied and the heavy schedules associated with the complicated planning of experiments in supine positions.

## Statistical Methods

Variables were analyzed by a two-way multiple ANOVA with time as the repeated measure (ambulatory versus bed rest) and group (control versus exercise) as main effect. RMR was analyzed by a repeated measures (ambulatory versus bed rest) analysis of covariance with FFM as the covariate and group (control versus exercise) as main effect. When necessary, within group comparisons were further evaluated by a paired t-test. Due to the low sample size inherent in such a study, all statistical results were confirmed by the Wilcoxon sign-rank. The Pearson product-moment correlation coefficient was used to investigate the relationships between relevant variables. All statistics were performed using JMP version 5.1.1 (SAS Institute, North Carolina, United States), and reported values are means ± SD (unless otherwise stated), with $p < 0.05$ considered statistically significant.

# RESULTS

## Participant Flow

The participant flow is represented in Figure 1.

## Recruitment

Recruitment of half the participants started in January 2001, and the experiment lasted from August 2001 to December 2001. The recruitment of the other half of subjects started in June 2001, and the experiment lasted from March 2002 to July 2002.

## Baseline Data

Table 1 shows the baseline characteristics of the participant groups. No major differences in these characteristics were noted at baseline.

## Numbers Analyzed

Nine participants in the control group and nine in the exercise group underwent the intervention. One subject of the exercise group was excluded from any procedures requiring ingestion or injection of products because of marginally high hepatic transaminase levels in the middle of the bed-rest period. This participant was used as the background subject given the small expected changes in background isotope enrichment over the time course of the study. One participant from the exercise group failed to comply with the training protocol due to a knee injury not reported during the selection process. He was also excluded from the analyses.

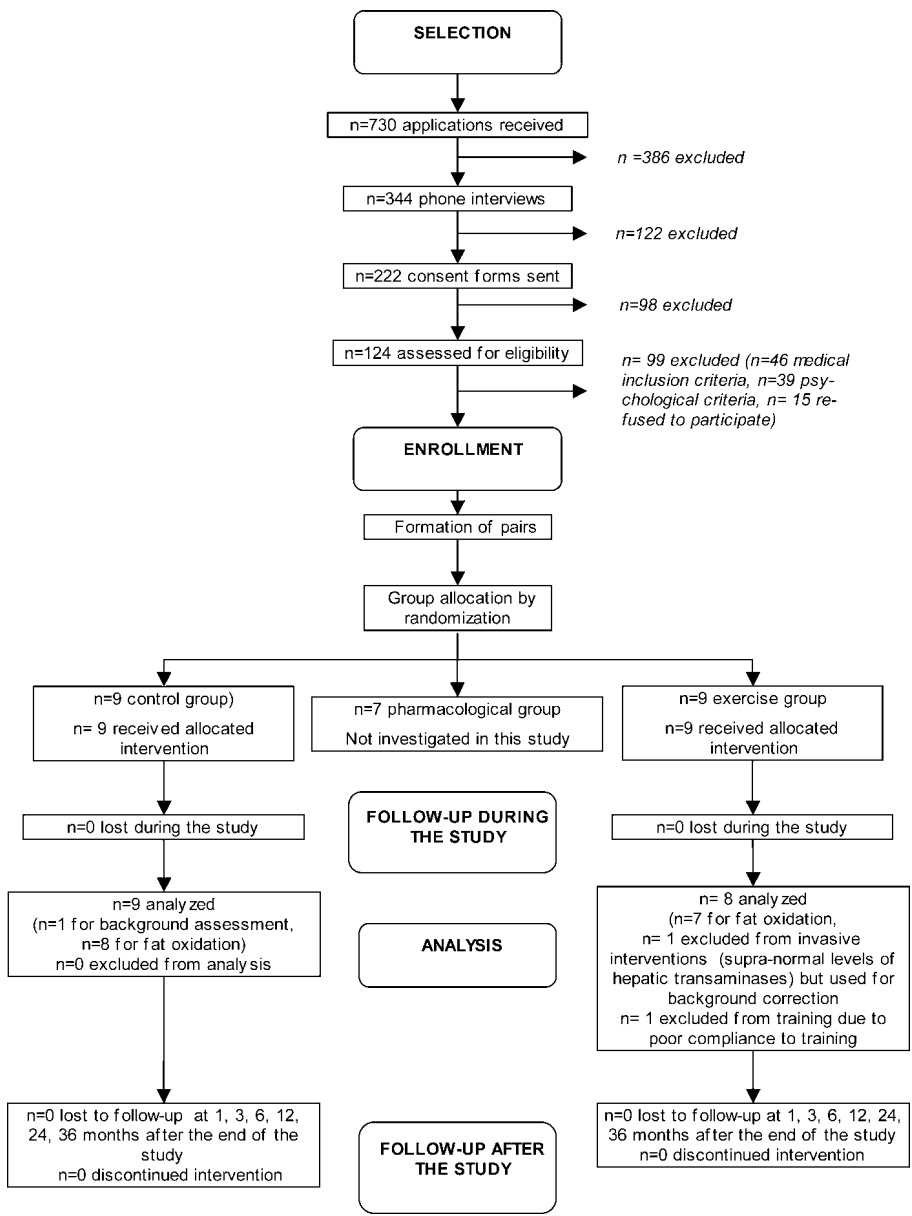

**Figure 1.** Flow Chart Showing the Progress of Participants in the Study
DOI: 10.1371/journal.pctr.0010027.g001

## Outcomes and Estimation

**Primary outcomes.** We did not observe any differences between the randomized groups for the primary outcomes. At the end of the bed rest, the mean difference between groups was 5.6% (95% CI, −3.3% to 14.5%; $p = 0.20$) for oleate cumulative recovery and −0.2% (95% CI, −4.1% to 3.6%; $p = 0.89$) for palmitate cumulative recovery at 7 h postdose. Within-subject changes were, however, noted.

The 7 h postdose cumulative recovery of palmitate (Figure 2) decreased after bed rest. In the control group, palmitate recovery was 22.7% ± 5.1% during the ambulatory period and 15.1% ± 3.9% during the bed-rest period (mean

**Table 1.** Characteristics of the Participants

| Variable | Unit | Control Group | Exercise Group |
|---|---|---|---|
| n | | 9 | 9 |
| Age | years | 32 (4) | 33 (5) |
| Height | m | 1.73 (0.03) | 1.75 (0.05) |
| Body mass | kg | 70.9 (6.1) | 70.6 (5.8) |
| Fat mass | % | 18.4 (4.7) | 15.6 (3.6) |
| Body mass index | kg/m$^2$ | 23.6 (1.9) | 23.0 (2.6) |

Values in parentheses indicate SD.
DOI: 10.1371/journal.pctr.0010027.t001

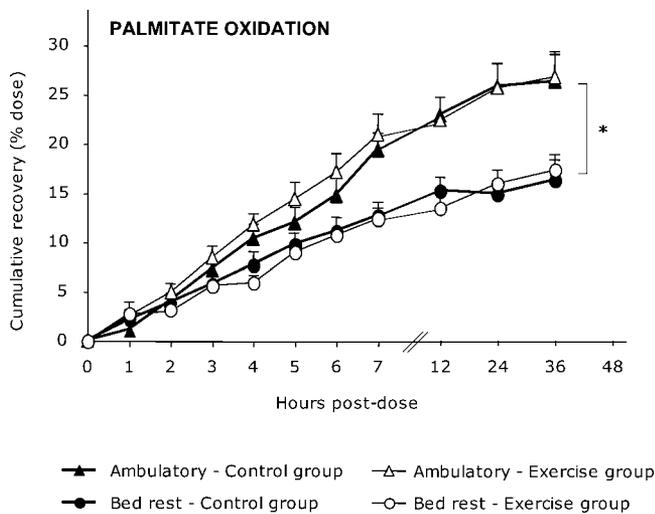

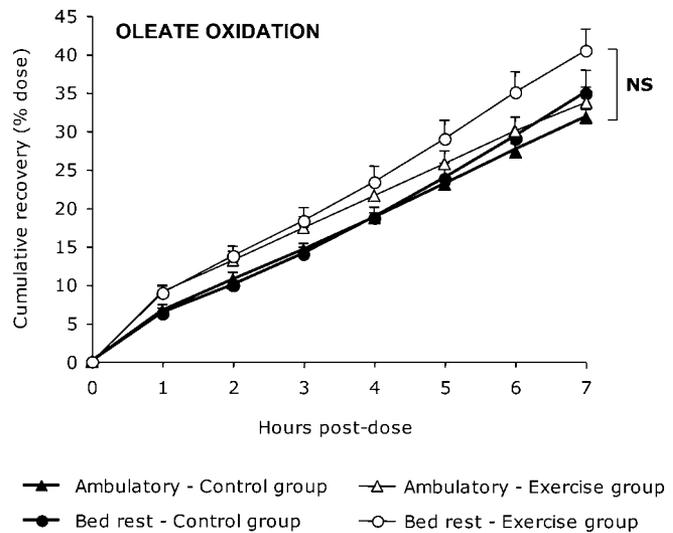

**Figure 2.** Hourly Cumulative Percent Recovery of [$d_{31}$]Palmitate (16:0) Before and After Bed Rest

This experiment included control ($n = 8$) and exercise ($n = 7$) groups. No between-group difference was noted after bed rest. *$p < 0.05$ compared to ambulatory period.
DOI: 10.1371/journal.pctr.0010027.g002

**Figure 3.** Hourly Cumulative Percent Recovery of [1-$^{13}$C]Oleate (18:1) Before and After Bed Rest

This experiment included control ($n = 8$) and exercise ($n = 7$) groups. Values are corrected by the acetate correction factor. No effects of exercise training or bed rest were noted (NS, not significant).
DOI: 10.1371/journal.pctr.0010027.g003

difference −6.7%; 95% CI, −10.5% to −3.0%; $p = 0.004$). In the exercise group, palmitate recovery was 22.3% ± 6.2% and 13.4% ± 4.7% during the ambulatory and bed-rest periods, respectively (mean difference −8.5%; 95% CI, −14.1% to −3.0%; $p = 0.01$). These changes were maintained for 36 h postdose. Cumulative recovery of palmitate of the control group at 36 h postdose was 28.5% ± 7.6% during the ambulatory period and 17.5% ± 5.6% during bed rest (mean difference −11.0%; 95% CI, −19.0% to −2.9%; $p = 0.01$). In the exercise group, oxidation dropped from 28.5% ± 7.0% to 17.2% ± 3.3% after bed rest (mean difference −11.3%; 95% CI, −18.4% to −4.2%; $p = 0.008$). Taken together, bed rest resulted in a −8.2% decrease in palmitate oxidation at 7 h postdose (95% CI, −10.4% to −4.7; $p < 0.0001$) and −11.1% at 36 h postdose (95% CI, −15.8% to −6.5%; $p < 0.0001$).

In contrast, no bed rest effects were noted on the 7 h postdose cumulative recovery of oleate (Figure 3). In the control group, oleate recovery was 31.7% ± 3.2% during the ambulatory period and 34.9% ± 8.6% during the bed rest (mean difference 3.2%; 95% CI, −4.2% to 10.5%; $p = 0.34$). In the exercise group, oleate recovery was 35.4% ± 6.8% and 39.4% ± 7.4% during the ambulatory and bed-rest periods, respectively (mean difference 6.8%; 95% CI, −1.2% to 14.7%; $p = 0.08$).

All results were confirmed by the nonparametric Wilcoxon sign-rank test ($p$-value not detailed).

**Secondary outcomes.** Energy intake via diet was 10.2 ± 0.7 MJ/d during the control period for both groups. By design, energy intake was decreased during the bed-rest period in the control (8.7 ± 0.7 MJ/d) and exercise (9.4 ± 0.6 MJ/d) groups to match calculated requirements. The between-group difference represents the estimated energy cost of exercise. For both groups, the macronutrient composition of the diet was 54% ± 1% carbohydrate, 29% ± 1% fat, and 16% ± 1% protein during the control period. During the bed-rest period the diet composition was 51% ± 2% carbohydrate, 31± 2% fat, and 17% ± 1% protein.

We observed a significant bed-rest effect on body mass and FFM, but the presence of a bed rest × group interaction indicated that the effect of bed rest differed between the control and exercise groups (Table 2). After three months of inactivity, body mass of the control group changed by −2.9 kg (95% CI, −5.1 to −0.6 kg), whereas no changes were noted in the exercise group (0.02 kg; 95% CI, −1.2 to 1.2 kg). The loss in the control group was essentially accounted for by the −2.4 kg change in FFM (95% CI, −3.9 to −0.9 kg) and no significant change in fat mass (−0.4 kg; 95% CI, −2.0 to 1.1 kg). The loss in FFM was confirmed by a Wilcoxon rank-sign test ($p = 0.008$). In the exercise group, the changes were not significant for FFM (−0.5 kg; 95% CI, −1.5 to 0.5 kg) and FM (0.5 kg; 95% CI, −0.5 to 1.5 kg). Of note, although FM showed no significant change over the time course of the bed rest and indicated that the control group averaged a 200 J/d negative energy balance and the exercise group a 300 J/d positive energy balance, between-participant variability in fat balance was large for both groups, ranging from −3.4 to +2.1 kg. These changes correlated with both energy (Pearson $r = 0.74$, $p < 0.01$) and fat intake (Pearson $r = 0.63$, $p < 0.05$) during the bed-rest period.

The fasting hormone and metabolite response to bed rest were not different in the control and exercise groups (Table 3). Except for glycerol and glucose, which remained unchanged, bed rest induced an increase in fasting β-hydroxybutyrate (30%), leptin (49%), insulin (132%), free fatty acids (17%), triglycerides (TGs) (52%), and the insulin to glucose ratio (132%). Leptin correlated strongly with FM during both the ambulatory and the bed-rest periods (Pearson $r = 0.68$; $p < 0.01$, and $r = 0.83$; $p < 0.0001$, respectively).

As expected, FFM was a determinant of RMR in both groups during both the ambulatory and bed-rest periods (Pearson $r = 0.65$; $p < 0.01$, and $r = 0.59$; $p < 0.01$, respectively). We observed a significant bed rest × group interaction for RMR (Table 4). Indeed, although RMR decreased by 8% after bed

**Table 2.** Body Mass and Composition

| Variable | Unit | Control Group | | Exercise Group | | ANOVA | | |
| --- | --- | --- | --- | --- | --- | --- | --- | --- |
| | | Ambulatory Period | Bed-Rest Period | Ambulatory Period | Bed-Rest Period | Bed Rest | Group | Bed Rest × Group |
| n | | 9 | 9 | 8 | 8 | | | |
| Body mass | kg | 70.9 (6.1) | 68.0 (4.6) | 69.0 (3.4) | 69.0 (4.2) | 0.03 | 0.84 | 0.02 |
| Fat mass | kg | 13.2 (4.1) | 12.8 (3.0) | 10.5 (2.8) | 11.0 (3.4) | 0.94 | 0.18 | 0.28 |
| Fat-free mass | kg | 57.6 (4.2) | 55.2 (2.8) | 58.5 (2.3) | 58.0 (1.6) | 0.003 | 0.21 | 0.03 |

Values are means (SD).
DOI: 10.1371/journal.pctr.0010027.t002

rest in the control group, no change was noted in the exercise group. These results were similar after adjustment for FFM. The average change for the control group was −630 kJ (95% CI, −929 to −332 kJ; Wilcoxon sign-rank test, $p = 0.008$) and −4 kJ (95% CI, −242 to 234 kJ, Wilcoxon test $p = 0.94$) for the exercise group. This suggests that the bed rest-induced decrease in RMR was independent of the active metabolic mass changes.

A similar shift in fasting substrate use was noted in both groups, either expressed in g/h or mg/kg FFM/h. The results per group are indicated in Table 4, but for clarity both groups have been combined in the following section. In mg/kg, FFM/h fasting glucose oxidation increased by 17% (23.1 mg/kg FFM/h; 95% CI, 11.7 to 34.5 mg/kg FFM/h; Wilcoxon sign-rank test, $p = 0.002$), and lipid oxidation decreased by 27% (−8.9 mg/kg FFM/h; 95% CI, −14.8 to −2.9 mg/kg FFM/h; Wilcoxon sign-rank test, $p = 0.01$). As a result, the NPRQ significantly increased by 4% (0.033; 95% CI, 0.054 to 0.013). Fasting lipid oxidation in g/min correlates with FM during both the ambulatory and bed-rest periods (Pearson $r = 0.68$; $p < 0.01$, and $r = 0.63$; $p < 0.05$, respectively). The 7 h postmeal cumulative substrate use expressed in g or mg/kg FFM showed the same pattern of response (Table 4). No group effect emerged, suggesting that resistance exercise training had no effect on the partitioning of postprandial substrate use. Expressed in mg/kg FFM and cumulative for both groups, postprandial glucose oxidation increased by 124 mg/kg FFM (95% CI, 28 to 220 mg/kg FFM; Wilcoxon sign-rank test $p = 0.005$) and lipid oxidation decreased by 36% (−53 mg/kg FFM; 95% CI, −86 to −21 mg/kg FFM; Wilcoxon sign-rank test $p = 0.003$). Lipogenesis, as measured by indirect calorimetry, increased by 129% (0.9 g; 95% CI, 0.2 to 1.6 g) but remained low in absolute numbers. Fasting and postprandial protein oxidation remained unaffected by bed rest, as did diet-induced thermogenesis (DIT).

### Ancillary Analyses
No associations were noted between the 7 h cumulative oxidation of oleate and various metabolic variables during the ambulatory or bed-rest periods (Table 5). Although the ambulatory 7 h cumulative palmitate oxidation showed the same results, strong associations were noted during bed rest. Palmitate oxidation correlates positively with fasting lipid oxidation and thus negatively with NPRQ. A positive relationship was also noted between palmitate oxidation and total fat oxidation over the 7 h postprandial phase (Table 5).

### Adverse Events
No adverse effects from the intervention were noted, and the participants tolerated all the experiments well.

## DISCUSSION
### Interpretation
Although the resistance exercise training program mitigated the physical inactivity-induced muscle atrophy (0.5 ± 1.2 kg lost in the exercise group versus 2.4 ± 1.9 kg lost in the control group), the substrate oxidation rates, the hyperinsulinemia, and the hyperlipidemia induced by bed rest were not restored by exercise to their basal values. Three points

**Table 3.** Fasting Hormones and Metabolites

| Variable | Unit | Control Group | | Exercise Group | | ANOVA | | |
| --- | --- | --- | --- | --- | --- | --- | --- | --- |
| | | Ambulatory Period | Bed-Rest Period | Ambulatory Period | Bed-Rest Period | Bed Rest | Group | Bed Rest × Group |
| n | | 8 | 8 | 7 | 7 | | | |
| Leptin | ng/ml | 1.7 (1.3) | 2.5 (1.4) | 2.0 (1.0) | 3.0 (2.1) | 0.04 | 0.53 | 0.80 |
| Insulin | mU/l | 25.8 (10.8) | 55.8 (23.2) | 20.6 (6.9) | 51.9 (10.2) | <0.0001 | 0.37 | 0.80 |
| Glucose | mmol/l | 4.9 (0.5) | 4.7 (0.2) | 4.6 (0.3) | 4.9 (0.3) | 0.61 | 0.78 | 0.06 |
| Insulin/glucose | | 5.3 (2.1) | 11.9 (4.7) | 4.5 (1.5) | 10.8 (2.7) | <0.0001 | 0.36 | 0.97 |
| β-hydroxybutyrate | μmol/l | 46.5 (17.3) | 58.0 (30.1) | 42.5 (8.9) | 57.6 (22.7) | 0.04 | 0.72 | 0.64 |
| Glycerol | μmol/l | 36.6 (13.8) | 39.6 (13.6) | 34.1 (12.7) | 39.9 (15.6) | 0.40 | 0.92 | 0.86 |
| Free fatty acids | μmol/l | 232.5 (138.1) | 275.5 (112.5) | 270.1 (56.3) | 311.3 (80.0) | 0.03 | 0.53 | 0.86 |
| Triglycerides | μmol/l | 897.0 (463.4) | 1341.5 (740.1) | 835.8 (390.6) | 1294.0 (966.7) | 0.02 | 0.89 | 0.97 |

Values are means (SD).
DOI: 10.1371/journal.pctr.0010027.t003

**Table 4.** RMR and Fasting and Postprandial Substrate Oxidation

| Variable | Variable Subcategory | Units | Control Group | | Exercise Group | | MANOVA | | |
|---|---|---|---|---|---|---|---|---|---|
| | | | Ambulatory Period | Bed-Rest Period | Ambulatory Period | Bed-Rest Period | Bed Rest | Group | Bed Rest × Group |
| n | | | 8 | 8 | 7 | 7 | | | |
| RMR | | MJ/d | 6.4 (0.4) | 5.9 (0.3) | 6.4 (0.3) | 6.4 (0.5) | 0.06 | 0.10 | 0.02 |
| RMR$_{FFM}$ | | MJ/d | 6.4 (0.3) | 5.8 (0.3) | 6.3 (0.3) | 6.3 (0.3) | 0.002 | 0.20 | 0.002 |
| Fasting substrate oxidation | NPRQ | | 0.890 (0.063) | 0.926 (0.058) | 0.890 (0.030) | 0.920 (0.040) | 0.01 | 0,91 | 0.79 |
| | Glucose | g/h | 7.5 (2.0) | 9.1 (1.9) | 8.0 (1.6) | 9.1 (1.4) | 0.003 | 0.79 | 0.49 |
| | | mg/kg FFM/hr | 134.2 (40.5) | 160.4 (37.4) | 134.2 (20.2) | 153.8 (20.7) | 0.01 | 0.74 | 0.49 |
| | Lipids | g/h | 1.9 (1.1) | 1.2 (0.9) | 1.9 (0.4) | 1.5 (0.7) | 0.001 | 0.84 | 0.55 |
| | | mg/kg FFM/hr | 32.6 (19.0) | 21.7 (16.0) | 31.8 (7.8) | 25.2 (10.4) | 0.01 | 0.85 | 0.46 |
| | Protein | g/h | 3.6 (0.6) | 3.6 (0.0) | 4.0 (0.6) | 3.9 (0.8) | 0.93 | 0.23 | 0.86 |
| | | mg/kg FFM/hr | 63.2 (10.0) | 64.1 (11.8) | 66.9 (11.5) | 66.6 (15.7) | 0.94 | 0.57 | 0.86 |
| 7 h postmeal cumulative substrate oxidation | DIT | % | 9.2 (2.1) | 10.6 (2.3) | 9.2 (2.0) | 9.9 (1.8) | 0.27 | 0.55 | 0.76 |
| | Glucose | g | 78.4 (11.8) | 82.9 (13.5) | 80.3 (12.8) | 89.7 (9.7) | 0.02 | 0.46 | 0.36 |
| | | mg/kg FFM | 1395 (249) | 1477 (230) | 1384 (187) | 1520 (171) | 0.01 | 0.99 | 0.33 |
| | Lipids | g | 7.5 (6.6) | 4.5 (5.0) | 10.0 (3.4) | 6.8 (3.4) | 0.004 | 0.34 | 0.93 |
| | | mg/kg FFM | 132 (117) | 79 (89) | 168 (59) | 114 (56) | 0.005 | 0.40 | 0.97 |
| | Protein | g | 25.3 (3.0) | 28.7 (5.7) | 27.7 (5.5) | 25.6 3.8 | 0.65 | 0.85 | 0.09 |
| | | mg/kg FFM | 452 (56) | 519 (131) | 464 (81) | 434 (56) | 0.51 | 0.33 | 0.10 |
| | Lipogenesis | g | 1.1 (1.6) | 1.9 (1.8) | 0.3 (0.4) | 1.3 (1.6) | 0.02 | 0.31 | 0.77 |

Values are mean (SD).
RMR$_{FFM}$, resting metabolic rate adjusted for fat-free mass.
DOI: 10.1371/journal.pctr.0010027.t004

can explain this lack of effect. First, resistance exercise training was shown to induce a decrease in baseline TG concentrations by 19% and an increase in resting fat oxidation by 21% when exercise was performed within 16 h postexercise [20]. We attempted to determine the effects of training, not the acute effects of the exercise. Our tests took place 36 h after the last session of training, and the delay may have been too long for the effect of training to pertain. Second, the impairment in oxidative capacity and the increase in glycolytic capacity observed in both groups may be partly explained by the pattern of changes in muscle fibers. Although the resistance exercise program was effective for maintaining whole muscle size, physical inactivity was shown to induce a shift from slow oxidative (type I) to fast glycolytic (types IIa and IIa/IIb) fibers and an apparition of hybrid fibers (type I/IIa/IIb), for which oxidative patterns are still unknown [21]. A higher proportion of type IIb fibers in skeletal muscle tissue was observed in obese and diabetic participants [22].

**Table 5.** Pearson Correlation Coefficients for Dietary Fat Oxidation

| Variable | Unit | Oleate 7 h Postdose | | Palmitate 7 h Postdose | |
|---|---|---|---|---|---|
| | | Ambulatory | Bed Rest | Ambulatory | Bed Rest |
| Body mass | kg | 0.35 | −0.28 | −0.19 | 0.31 |
| FFM | kg | 0.19 | −0.11 | 0.08 | 0.17 |
| FM | kg | 0.31 | −0.26 | −0.42 | 0.30 |
| RMR | MJ/d | 0.18 | −0.05 | 0.06 | 0.10 |
| Fasting NPQR | | −0.02 | 0.41 | 0.29 | −0.74** |
| Fasting lipid oxidation | g/mn | 0.10 | −0.39 | −0.29 | 0.77*** |
| Fasting glucose oxidation | g/mn | 0.17 | 0.17 | 0.16 | −0.61* |
| Breakfast energy content | kcal | −0.27 | 0.10 | −0.15 | 0.11 |
| Breakfast lipid content | kcal | −0.04 | −0.16 | 0.22 | 0.44 |
| Breakfast glucose content | kcal | −0.33 | 0.20 | −0.06 | −0.04 |
| 7 h postmeal lipid oxidation | g | 0.24 | −0.15 | 0.15 | 0.74** |
| 7 h postmeal glucose oxidation | g | −0.23 | −0.08 | −0.15 | −0.38 |
| DIT | % | 0.36 | 0.14 | 0.03 | −0.41 |
| Insulin | mU/l | −0.03 | 0.05 | −0.15 | −0.07 |
| Glucose | mmol/l | −0.07 | 0.17 | −0.15 | −0.23 |
| Free fatty acids | μmol/l | 0.35 | −0.06 | 0.32 | 0.50 |
| Triglycerides | μmol/l | 0.01 | 0.36 | −0.47 | −0.37 |

* $p < 0.05$; ** $p < 0.01$; *** $p < 0.0001$
DOI: 10.1371/journal.pctr.0010027.t005

Finally, the modest energy expenditure induced by our resistance exercise training program (about 640 kJ per session) may not have been sufficient to fully protect the muscle fiber pattern and the related oxidative capacity. In support of that, an early bed-rest study suggested a relationship between the amount of energy expenditure and the degree of insulin resistance [23]. Overall, physical inactivity may induce a muscle fat oxidative capacity impairment that leads to a drop in fat oxidation rate and to insulin resistance. This study provides evidence that physical inactivity itself triggers physiopathological characteristics observed in obese and diabetic individuals.

The second finding of our investigation indicated that physical inactivity per se induces a preferential partitioning of dietary fat from oxidation toward storage. Studies on exercise training and detraining [24–26] may provide insights into putative mechanisms of this process, as the effects of exercise training on postprandial lipidemia have been well documented. The rise in dietary TG clearance after prior exercise is likely to be mediated in part by an up-regulation of lipoprotein lipase (LPL) activity. Exercise for 5–13 consecutive days increases LPL mRNA in skeletal muscle with no obvious changes in adipose tissue LPL mRNA [27]. Conversely, a two-week period of detraining in runners decreased LPL activity in skeletal muscle but increased LPL activity in adipose tissue [28]. Further investigations are therefore needed to better understand the fat oxidation regulation mechanisms according to physical activity level.

A notable finding of this study is that the longitudinal effects of physical inactivity on dietary fat oxidation depend on the nature of the fatty acid. Physical inactivity decreased saturated, but not monounsaturated, fatty acid oxidation. We observed that oleate recovery as a percentage was higher than that for palmitate, regardless of the testing period. This result is in agreement with previous studies showing that polyunsaturated and monounsaturated fatty acids are more oxidized in inactive persons than are saturated fatty acids [29]. Recent investigations in men have shown that dietary fat type substitution may affect energy metabolism and ultimately body mass. Piers and colleagues [30] have reported that the postprandial fat oxidation rate is higher, and carbohydrate oxidation rate is lower, after a monounsaturated fatty acid, rather than after saturated fatty acid meal. A further four-week dietary intervention on the same participants in which saturated fatty acids were replaced by monounsaturated fatty acids induced a significant loss of body weight and fat mass without any change in total energy or fat intake [31].

Recently, Kien et al. [32] fed healthy young adults either with a high palmitic acid diet or with a high oleic acid diet and showed a reduction in postprandial lipid oxidation and daily energy expenditure following the high-palmitate diet and no changes after the high-oleate one. In our study, the environmental intervention did not involve energy or fat balance, but rather involved physical activity level. Previous studies clearly showed that dietary fat oxidation is unrelated to global fat oxidation, and we confirmed this result in the control period for both oleate and palmitate. Under bed-rest conditions, however, although oleate oxidation seemed to remain independent of global substrate use, palmitate oxidation was strongly correlated with fasting NPRQ as well as with postprandial lipid oxidation. This decrease in palmitate oxidation induced by physical inactivity may be, at least in part, explained by the shift from red to white muscles. Palmitate transport into giant sarcolemmal vesicles and hence, palmitate oxidation, are significantly greater in red oxidative muscles than in white glycolytic muscles [33]. Additionally, the muscle fiber type muscle proportion may also explain the inefficiency of resistance exercise training on palmitate oxidation. The reason for the unchanging oleate oxidation are unknown. A similar differential fat oxidation pattern was noted in type 2 diabetes, a disease that shares numerous features with obesity, including a decreased capacity to use fat as fuel. A study in cultured human myotubes from diabetic participants reported a higher oxidation of oleate compared to palmitate. Whereas oleate preferentially appeared into the intramyocellular free-fatty acid pool, palmitate was incorporated into intramuscular TG [34]. The pathways at play require further investigation.

## Overall Evidence and Generalizability

Although epidemiological studies have demonstrated relationships between sedentary behaviors and morbidity [6], most of our knowledge on mechanism comes from studies on the beneficial effects of adding exercise training. This is also the case for our understanding of the regulation of fat oxidation [20,35–37]. One study showed that occasional inactivity lowered fat tolerance to high-fat diet, leading potentially to weight gain in the long term [38]. Although impaired fat oxidation is considered causative in obesity, our knowledge on the chronic effects of sedentary behaviors on oxidative balance, independently of energy balance disturbances, is still weak. The present study is one of the first longitudinal studies to show that physical inactivity per se impairs dietary fat oxidation. We used a unique model of physical inactivity mainly utilized by space agencies for microgravity simulation: long-term bed rest, which decreases physical activity level from 1.7 to 1.2 [18]. Although we acknowledge that the physical inactivity induced by strict bed rest is severe and might not represent the level achieved in the general population, such a study design clearly helps to clarify mechanisms and possible levers of action on which countermeasures may be tested.

With the understanding that inactivity is no longer applicable at a certain volume of exercise, the key is to find the minimal amount of exercise that will restore dietary fat oxidation. This notion of minimal activity level is currently a matter of great debate in the literature. We failed to show that high-intensity low-volume resistance exercise training could mitigate the changes induced by bed rest on fat oxidation. The low level of energy expenditure attained during resistance exercise might explain those results. Another trial has been completed that will provide evidence on the effect of training at higher expenditure (SB, personal communication).

## Study Limitations

Several limitations have to be considered in our study. First, we acknowledge the fact that the sample size is low. Long-term bed rest of such duration represent a clear challenge to implement, and large sample sizes are not realistic from both practical and economic points of view. The involvement of different teams testing different study hypotheses on the same group of participants limits the a priori calculation of power.

Second, a modest but significant 2 g per meal increase in fat intake occurred during the bed-rest period despite a strict monitoring of nutritional conditions. Such a change in percentage fat intake is, however, unlikely to be responsible for the reduction in postprandial fat oxidation, as a major change in fat intake (20 to 50 g) has only modest effects on dietary fat oxidation [39]. We also failed to accurately stabilize fat mass and interindividual variations were large and ranged from −3.4 to +2.1 kg. Nevertheless, energy imbalance is unlikely to explain the results because average fat mass remained stable (0.3 ± 1.4 kg).

Last, we observed a trend toward higher oleate oxidation in the exercise group after bed rest ($p = 0.08$). We can not exclude the possibility that time points over the 7 h postdose might result in a significant effect of training on oleate, as observed by Votruba et al. [40]. This was not possible due to experiment constraints in this bed-rest study, but is currently being tested in women submitted to two months of bed rest.

## Conclusion

The present study shows that physical inactivity plays a key role in the portioning of saturated fat to oxidation, independently of its effects on energy balance, and results in a metabolic state comparable to that observed in obesity. Further investigations are necessary to better understand the underlying mechanisms and the parameters of an adequate exercise training program (in terms of frequency, intensity, duration, and type of exercise). Our study suggests that the Mediterranean diet (that is, one low in saturated fats, high in monounsaturated fats) would be helpful if promoted in sedentary populations and in recumbent patients, two groups at risk for weight gain.

...................................................................

## SUPPORTING INFORMATION

**CONSORT Checklist**
Found at DOI: 10.1371/journal.pctr.0010027.sd001 (54 KB DOC).

**Trial Protocol Part A**
Found at DOI: 10.1371/journal.pctr.0010027.sd002 (467 KB DOC).

**Trial Protocol Part B**
Found at DOI: 10.1371/journal.pctr.0010027.sd003 (1.1 MB DOC).

**Protocol S1.** Detailed Methods
Found at DOI: 10.1371/journal.pctr.0010027.sd004 (67 KB DOC).

## ACKNOWLEDGMENTS

The authors are indebted to the entire staff of the Institute of Space Medicine for the outstanding organization of the bed rest supported by CNES, ESA, and NASDA; special thanks are due to Alain Maillet, Marie-Pierre Bareille, Pascale Vasseur, and Brigitte Limouzin. We also acknowledge the outstanding support of the medical team at MEDES, especially Magali Cheylusse, Pascale Cabrole, and Maud Julien. We thank Corine Louche-Pelissier for performing hormone and metabolite assays. The most credit, however, must be given to the participants of the study.

## Author Contributions

GGK, HO, CG, and SB designed the study. SN, ML, TS, and SB analyzed the data. DAS, SN, GGK, MD, and HO collected data or did experiments. AB wrote the first draft of the paper. AB, DAS, SN, ML, YLM, CG, and SB contributed to the writing of the paper.